\documentclass[%
reprint,
superscriptaddress,
aps,
pra,
]{revtex4-2}

\usepackage[pdftex]{graphicx, xcolor}
\usepackage{float}
\usepackage{dcolumn}
\usepackage{amsmath,amssymb,bm}
\usepackage{braket}
\usepackage{url}
\usepackage{here}
\usepackage{ulem}

\newcommand{\bd}{{\bf d}}

\newcommand{\br}{{\bf r}}

\newcommand{\bE}{{\bf E}}
\newcommand{\bG}{{\bf G}}

\newcommand{\bI}{{\bf I}}

\newcommand{\nm}{\nonumber}

\renewcommand{\Re}{\operatorname{Re}}
\renewcommand{\Im}{\operatorname{Im}}

\begin{document}


\title{
Spectral design principles for local-excitation retention in impurity-assisted atomic arrays
}

\author{Junpei Oba}
\email[Contact author: ]{junpei-oba@mosk.tytlabs.co.jp}
\affiliation{Toyota Central R\&D Labs., Inc., 41-1 Yokomichi, Nagakute, Aichi 480-1192, Japan}

\date{\today}

\begin{abstract}
Enhanced local-excitation retention in atomic arrays allows to exploit cooperative radiative effects to suppress emission and prolong excited-state lifetimes. We consider an impurity-assisted setting involving a single storage atom being initially excited and study the survival of local excitation under neither write nor retrieval fields. Because the corresponding dynamics can involve multiple interfering collective modes, the survival dynamics cannot determined from the smallest collective decay rate alone. Thus, using a biorthogonal eigenmode decomposition of an effective non-Hermitian Hamiltonian, we show that the survival dynamics are jointly governed by the decay rates of the eigenmodes and their overlaps with the initial excitation. Large oscillations occur when multiple long-lived modes have comparable weights. Accordingly, we introduce a physically motivated spectral surrogate objective that favors both small weighted decay rates and an initial-state weight concentrated on a single subradiant mode. As a proof of principle of this spectral design, we apply the surrogate to constrained atom-position optimization under minimum-distance constraints and obtain nontrivial aperiodic configurations with enhanced local-excitation retention. Our findings unveil spectral design principles for local-excitation retention in impurity-assisted atomic arrays and provide a proof of principle for their inverse design.
\end{abstract}

\maketitle

\section{Introduction}
\label{sec:Introduction}

Light--matter interaction is pivotal in a wide range of phenomena and applications in quantum information processing, precision metrology, and nanophotonics~\cite{Novotny2012, Chang2018}.
Remarkably, in spontaneous emission, an initially excited atom relaxes to its ground state while emitting a photon, with a characteristic decay rate determined by the atomic transition.
In an atomic ensemble, however, radiative decay can be strongly altered by photon-mediated dipole--dipole interactions~\cite{Dicke1954, Lehmberg1970, Jen2025}.
These interactions include superradiance~\cite{Gross1982, Masson2022, Sierra2022, Rubies-Bigorda2023a}, which enhances emission, and subradiance~\cite{Bienaime2012, Garcia2017, Moreno-Cardoner2019, Rubies-Bigorda2023b}, which suppresses emission and prolongs excited-state lifetimes.
Because strong interactions require interatomic separations on the order of (or below) the optical wavelength, experimental access has remained limited, but recent progress has enabled the observation and control of cooperative emission in platforms such as 
microcavities, nanophotonic waveguides, and optical lattices~\cite{DeVoe1996, Bloch2008, Mlynek2014, Guerin2016, Solano2017, Jenkins2017, Rui2020, Zhou2024}.
These advances have motivated the use of collective radiative engineering for high-bandwidth quantum memories~\cite{Rastogi2022} and longer-lived quantum memories~\cite{Garcia2017, Bonanno2022}.
Remarkably, in photon storage~\cite{Facchinetti2016, Ballantine2021, Rubies-Bigorda2022}, a localized optical excitation is retained in an atomic system for a long time and then retrieved.
Hence, a designated storage atom can be coupled to a surrounding array that can host long-lived collective (subradiant) modes.

\begin{figure}[h]
 \centering
 \includegraphics[width=0.99\linewidth]{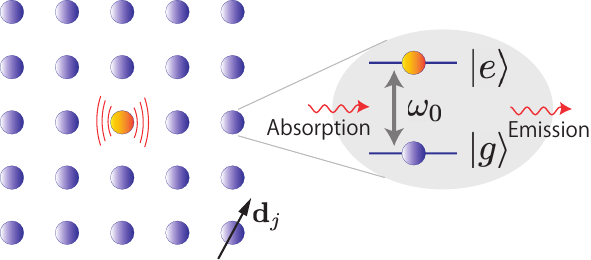}
  \caption{
  Conceptual diagram of considered system.}
\label{fig:setup}
\end{figure}

A simple cooperative setting is impurity-assisted storage, where a designated storage atom is initially excited and coupled to a surrounding atomic array (Fig.~\ref{fig:setup}).
In favorable geometries, local excitation hybridizes with collective modes of the array and acquires a long-lived subradiant component.
We focus on the survival probability of this initially prepared one-excitation state in the absence of explicit write and retrieval fields.
Instead of optimizing a complete write--store--read protocol, we identify spectral design principles for local-excitation retention in an impurity-assisted cooperative system.

Various advantageous geometries and control parameters have been explored for cooperative storage.
Buckley-Bonanno {\it et al.}~\cite{Bonanno2022} analyzed the minimum decay rates of periodic arrays by using an effective decay-rate model that enables rapid scans over the storage-atom position and detuning.
However, when the dynamics involve multiple collective modes, time dependence may be unsuitably described by a single exponential. 
Consequently, the performance depends on both the minimization of the decay rate and overlap strength between the initial excitation and relevant long-lived modes.
This multimode characteristic also hinders direct time-domain optimization.
A common objective, like maximizing the excitation probability at a selected final time, can depend sensitively on that choice under oscillating dynamics.

We seek a description that remains informative even under several collective modes contributing notably to the dynamics. 
To this end, we adopt an eigenmode-based viewpoint based on the spectrum of an effective non-Hermitian Hamiltonian.
In detail, biorthogonal decomposition~\cite{Brody2014} expresses the survival amplitude as a sum over collective modes with complex eigenvalues and mode-dependent overlaps with the initial state.
This representation allows us to identify the spectral features that control the dynamics.
As a result, the smallest collective decay rate alone is not sufficient to predict the performance, and pronounced oscillations appear when multiple long-lived modes carry comparable weights and interfere at frequencies that depend on the differences in the real parts of the eigenvalues.
We compare representative examples of geometries to extract relevant spectral mechanisms rather than establish a universal ranking of lattice classes.

As a proof of principle for this spectral viewpoint, we perform constrained atom-position optimization.
The objective function is designed as a physically motivated surrogate that favors both a long lifetime and concentration of the initial-state weight in a single dominant subradiant mode while directly imposing experimentally motivated constraints, such as a minimum interatomic distance.
Hence, nontrivial aperiodic configurations can be inversely designed with enhanced local-excitation retention from a simple input configuration.

This study provides three specific contributions. 1) We identify a spectral criterion for local-excitation retention in the multimode regime, where performance is jointly governed by modal decay rates and initial-state overlaps. 2) We introduce a physically motivated spectral surrogate that compactly captures these two requirements. 3) As a proof of principle, we embed this surrogate for constrained atom-position optimization and obtain aperiodic configurations with enhanced retention under different minimum-distance constraints.

The remainder of this paper is organized as follows. 
Section~II presents the model and biorthogonal eigenmode decomposition.
Section~III describes the computational details.
In Sec.~IV, we present and discuss an eigenanalysis of the subradiant dynamics and results of atom-position optimization.
Finally, we draw conclusions in Sec.~V.

\section{Model}
\label{sec:Model}

\subsection{Equations of motion for atomic degrees of freedom}

We consider a system of $N$ two-level atoms (ground state $\ket{g_j}$, excited state $\ket{e_j}$, transition frequency $\omega_0$, dipole moment $\bd_j$, and position $\br_j$) interacting with the vacuum electromagnetic field.
To describe the decay dynamics of a storage atom, we restrict the number of excitations to one and neglect any external input field.
The time evolution of the atomic degrees of freedom follows the Schr\"odinger equation~\cite{Bonanno2022}:
\begin{align}
\frac{d}{dt}\ket{\psi(t)} &= -i \hat{H}\ket{\psi(t)},
\label{eq:dot_bc}
\end{align}
where $\ket{\psi(t)}=\sum_j c_j(t) \ket{e_j} \bigotimes_{j'\neq j} \ket{g_{j'}}$ and
$\hat{H}=\sum_{jj'}\left(J_{jj'}-\frac{i}{2}\Gamma_{jj'}\right)\hat{\sigma}^{j}_{eg}\hat{\sigma}^{j'}_{ge}$.
Because transitions in which a photon emitted from an excited atom escapes the system correspond to dissipation regarding the atomic degrees of freedom, $\hat{H}$ is a non-Hermitian Hamiltonian.
We neglect reflective boundaries, periodic boundary conditions, or other mechanisms that would return photons to the system.
The probability that the $j$-th atom is in the excited state at time $t$ is given by $|c_j(t)|^2$.
Operator $\hat{\sigma}^{j}_{\alpha\beta}=\ket{\alpha_j}\!\bra{\beta_j}$ ($\alpha,\beta\in\{e,g\}$) acts on the $j$-th atom.
Coefficients $J_{jj'}$ and $\Gamma_{jj'}$ are given by
\begin{align}
J_{jj'} &= -\frac{3\pi \sqrt{\gamma_j \gamma_{j'}}}{\omega_0}\,
\hat{\bd}_j^\ast \cdot \Re\!\left[\bG_0(\br_{j} - \br_{j'},\omega_0)\right]\cdot \hat{\bd}_{j'}, \\
\Gamma_{jj'} &= \frac{6\pi \sqrt{\gamma_j \gamma_{j'}}}{\omega_0}\,
\hat{\bd}_j^\ast \cdot \Im\!\left[\bG_0(\br_{j} - \br_{j'},\omega_0)\right]\cdot \hat{\bd}_{j'},
\label{eq:Gamma}
\end{align}
which represent the cooperative Lamb shift and cooperative decay rate, respectively~\cite{Bonanno2022},
with $\gamma_j$ being the spontaneous emission rate of the $j$-th atom when it exists alone in vacuum, 
$\bG_0(\br_{j} - \br_{j'},\omega_0)$ being the Green's tensor of the three-dimensional vacuum electromagnetic field, and
$\hat{\bd}_j=\bd_j/|\bd_j|$ being the unit vector along the dipole moment of the $j$-th atom.
In addition, $\lambda_0$ denotes the wavelength of light corresponding to energy $\omega_0$.

\subsection{Eigendecomposition of non-Hermitian Hamiltonian}

Assume that the effective non-Hermitian Hamiltonian $\hat{H}$ is diagonalizable.
Thus, the right and left eigenstates form a complete biorthogonal basis upon appropriate normalization, such that the time-evolution operator can be expanded in terms of the eigenmodes.

Let $\ket{\psi^R_\ell}$ and $\ket{\psi^L_\ell}$ ($\ell=1,\dots,N$) be the right and left eigenstates, respectively.
These eigenstates are obtained as solutions of
\begin{align}
\hat{H}\ket{\psi^R_\ell} &= \kappa_\ell \ket{\psi^R_\ell}, \\
\hat{H}^\dag \ket{\psi^L_\ell} &= \kappa_\ell^\ast \ket{\psi^L_\ell}.
\end{align}
Then, time-evolution operator $e^{-i\hat{H}t}$ can be decomposed as~\cite{Brody2014}
\begin{align}
e^{-i\hat{H}t}
= \sum_\ell e^{-i\kappa_\ell t}\,
\frac{\ket{\psi^R_\ell}\!\bra{\psi^L_\ell}}{\braket{\psi^L_\ell|\psi^R_\ell}}.
\label{eq:expH}
\end{align}
The quantum state at time $t$ is given by
\begin{align}
\ket{\psi(t)} &= e^{-i\hat{H}t}\ket{\psi (0)} \nonumber\\
&= \sum_\ell e^{-i\kappa_\ell t}\,
\frac{\ket{\psi^R_\ell}\!\bra{\psi^L_\ell}}{\braket{\psi^L_\ell|\psi^R_\ell}}\ket{\psi (0)}.
\end{align}
Therefore, the probability that the system remains in the initial state at time $t$ is given by
\begin{align}
p_e(t) &= \left|\braket{\psi (0)|\psi(t)}\right|^2,
\label{eq:pe} \\
\braket{\psi (0)|\psi(t)}
&= \sum_\ell e^{-i\kappa_\ell t}\,
\frac{\braket{\psi (0)|\psi^R_\ell}\,\braket{\psi^L_\ell|\psi (0)}}{\braket{\psi^L_\ell|\psi^R_\ell}} \nm\\
&\equiv \sum_\ell w_\ell e^{-i\kappa_\ell t},
\label{eq:pe_inside}
\end{align}
where weight $w_\ell$ represents the overlap between the initial state and eigenstates.
Weight $w_\ell$ is generally complex in non-Hermitian systems. 
For a normalized initial state, $\sum_\ell w_\ell = 1$ follows from Eq.~\eqref{eq:pe_inside} at $t=0$.
We also define the decay rate of the $\ell$-th eigenmode as $\Gamma_\ell \equiv -2\,\mathrm{Im}\,\kappa_\ell$.

Numerical calculations show that the standard biorthogonal eigendecomposition is numerically well behaved for all structures considered in this study. The normalized residuals of the computed right and left eigenstates with order $10^{-15}$ are given by
\begin{align}
&\frac{\|\hat{H}\ket{\psi_\ell^{R}}-\kappa_\ell\ket{\psi_\ell^{R}}\|}{\|\hat{H}\|\,\|\ket{\psi_\ell^{R}}\|}, \\
&\frac{\|\hat{H}^\dagger\ket{\psi_\ell^{L}}-\kappa_\ell^{*}\ket{\psi_\ell^{L}}\|}{\|\hat{H}\|\,\|\ket{\psi_\ell^{L}}\|}.
\end{align}
Hence, the right and left eigenstates are satisfied to numerical precision for the considered structures, and the eigenmode expansion used in Eqs.~\eqref{eq:expH}--\eqref{eq:pe} is numerically accurate.
Note that $\|\cdot\|$ represents the Frobenius and Euclidean norms for matrices and vectors, respectively.

\section{Implementation details}
\label{sec:Implementation_details}

\begin{figure*}[ht]
 \centering
 \includegraphics[width=0.9\linewidth]{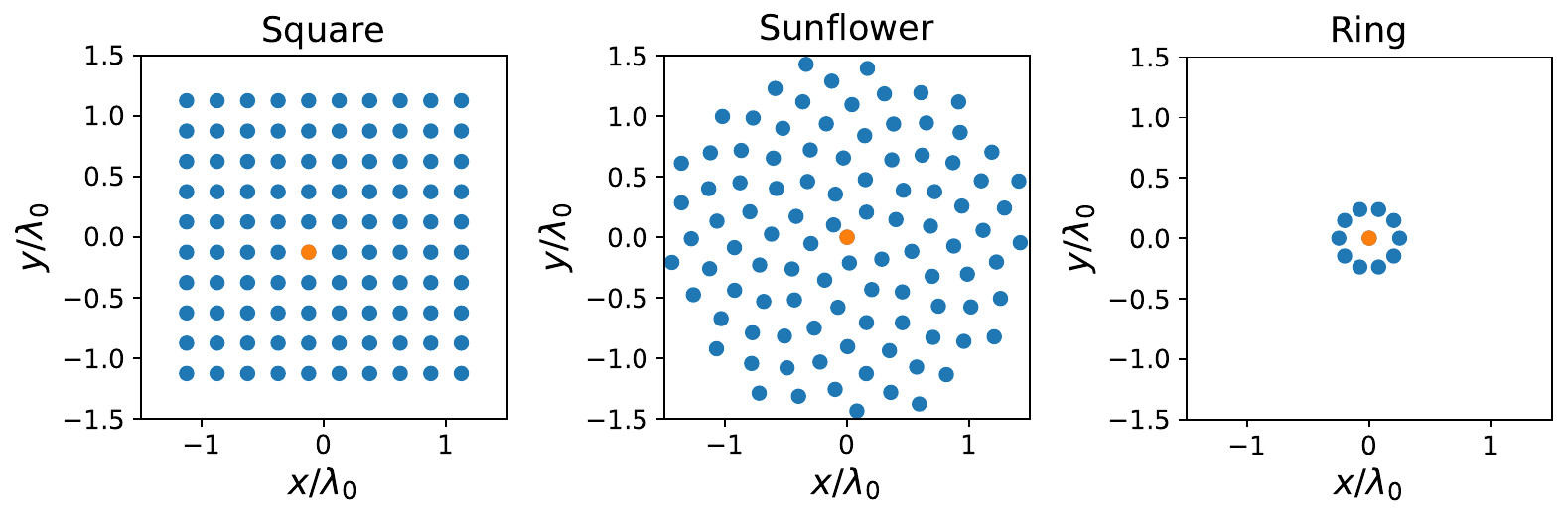}
 \caption{
 Atomic configurations. The orange dot indicates the storage atom prepared in the excited state at $t=0$,
 and the blue dots indicate atoms initially in the ground state.
 The parameters are set to $n=10$ and $a=0.25$.
 For $n=10$, the square and sunflower configurations contain 100 surrounding atoms,
 whereas the ring configuration contains 10 surrounding atoms.
 These geometries are used as representative examples rather than as a controlled benchmark at a fixed atom count.}
 \label{fig:structure}
\end{figure*}

We set $\hbar=c=1$ and $\omega_0=2\pi$.
All atoms, including the storage atom, are assumed to have the same spontaneous emission rate, $\gamma_j=\gamma_0$.
In addition, we assume that all atoms share dipole orientation $\hat{\bd}$.
For the spatial configurations of the atomic ensemble, we employ a square lattice as a periodic structure and a sunflower pattern and ring as aperiodic structures (Fig.~\ref{fig:structure}).
Each structure is specified by atom count $n$ and distance $a$.
For the square lattice, atoms are placed on an $n\times n$ grid with nearest-neighbor spacing $a\lambda_0$.
For the sunflower pattern, $n\times n$ atoms are arranged such that the polar coordinates of the $j$-th atom are $(r_j, \theta_j) = (\sqrt{j}r_c, j \pi (3 - \sqrt{5}))$, with $r_j$ being a constant that ensures average nearest-neighbor distance $a\lambda_0$.
For the ring configuration, $n$ atoms are placed on a circle with radius $a\lambda_0$.
The storage atom is assigned to the central atom.
Let $j^\ast$ denote the index of the storage atom. The initial state is expressed as $\ket{\psi (0)} = \ket{e_{j^\ast}} \bigotimes_{j \neq j^\ast} \ket{g_j}$.
For the current initial condition, the survival probability defined in Eq.~\eqref{eq:pe} coincides with the excitation probability of the storage atom (i.e., $p_e (t) = |c_{j^\ast} (t)|^2$), and we use this quantity as the performance measure.

\section{Results and discussion}
\label{sec:Results_and_discussion}

\subsection{Subradiant dynamics and spectral criteria}

\begin{figure*}[ht]
 \centering
 \includegraphics[width=0.99\linewidth]{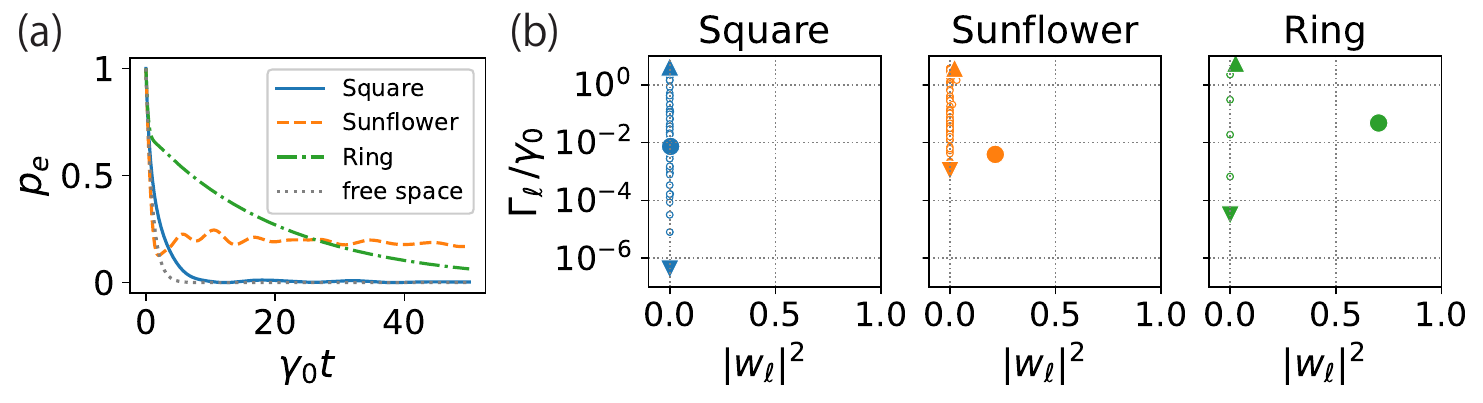}
 \caption{
 (a) Decay dynamics for different atomic configurations. The gray dotted line indicates the free-space decay dynamics of a single atom.
 (b) Scatter plots of decay rate and weight per eigenmode.
 From left to right: square lattice, sunflower, and ring configurations.
 The three representative geometries illustrate spectral mechanisms, but they are not intended to establish a universal ranking of geometry classes.
 The upward and downward triangles indicate the eigenmodes with the maximum and minimum decay rates, respectively.
 The filled circles indicate the eigenmode with the largest weight.
 The parameters settings are $n=10$, $a=0.25$, and $\hat{\bd}=(1,i,0)/\sqrt{2}$.}
 \label{fig:Gamma_weight}
\end{figure*}

\begin{figure}[ht]
 \centering
 \includegraphics[width=0.9\linewidth]{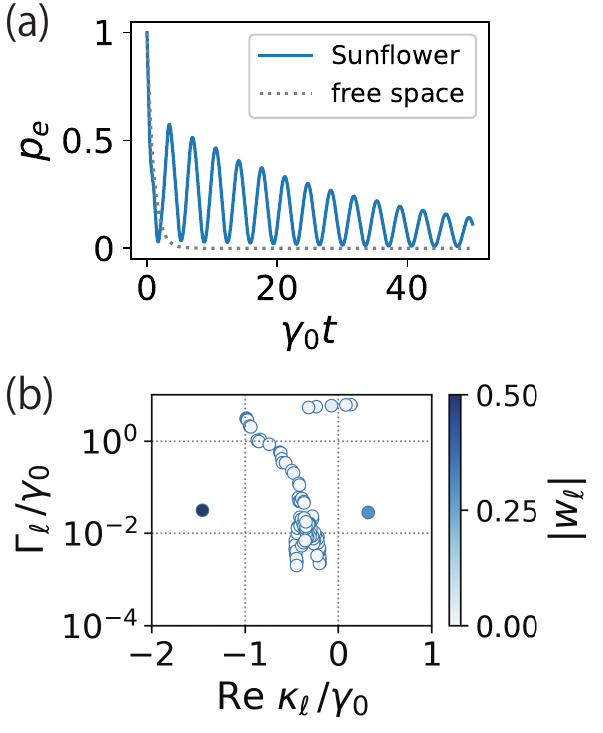}
 \caption{
 (a) Example of pronounced oscillations in excitation probability for sunflower configuration.
 The parameters settings are $n=10$, $a=0.2$, and $\hat{\bd}=(0,0,1)$.
 (b) Scatter plot of decay rate and real part of eigenvalue per eigenmode.
 The color intensity of each marker indicates the weight.
 Two eigenmodes with large weights and comparably small decay rates are observed.}
 \label{fig:Gamma_Re}
\end{figure}

We first compare the three representative geometries depicted in Fig.~\ref{fig:structure} to identify spectral quantities that are relevant to the excitation-survival dynamics before structure optimization.
Because these three examples are not matched in the total atom count, density, or spatial extent, the comparison should not be interpreted as a controlled ranking of geometry classes.
Instead, they are intended to extract qualitative spectral mechanisms that are relevant to subsequent optimization.

Figure~\ref{fig:Gamma_weight}(a) shows the decay dynamics for the three representative examples.
Figure~\ref{fig:Gamma_weight}(b) shows scatter plots of the decay rate and weight per eigenmode.
The minimum decay rate alone does not allow one to predict the observed dynamics. Among the examples, the square lattice contains an eigenmode with the smallest decay rate, but its excitation survival is poorer than those for the sunflower and ring geometries.
Hence, the decay dynamics of the storage atom are governed not only by small decay rates but also by the overlapping strength between the initial excitation and corresponding eigenmodes.

For the square lattice, the modal weights are broadly distributed, and no single long-lived mode dominates the initial state.
In the sunflower and ring geometries, a dominant mode with a much larger weight appears per geometry.
From these examples, this difference accounts for the trends shown in Fig.~\ref{fig:Gamma_weight} more directly than the minimum decay rate alone. The sunflower geometry exhibits a longer lifetime because its dominant mode has a smaller decay rate than that of the ring geometry, which in turn shows a higher early excitation probability because its dominant mode carries a larger weight.
Therefore, Fig.~\ref{fig:Gamma_weight} motivates the design principle that both decay rates and overlaps must be considered simultaneously.

Oscillations appear in the excitation probability for the sunflower geometry, and we attempt to interpret such dynamics via eigendecomposition.
Figure~\ref{fig:Gamma_Re}(a) shows an example of decay dynamics with pronounced oscillations.
These oscillations arise from repeated transitions, in which a photon propagates to the surrounding atoms and subsequently returns to the storage atom.
Large oscillations, as shown in Fig.~\ref{fig:Gamma_Re}(a), are undesirable in the present setting for local-excitation retention because the excitation probability of the storage atom depends strongly on the measurement time.

Analytically, consider Eq.~\eqref{eq:pe}, in which only two eigenmodes have large weights. Then,
\begin{align}
p_e(t) &= \left| w_1 e^{-i \kappa_1 t} + w_2 e^{-i \kappa_2 t} \right|^2 \nonumber\\
&= |w_1|^2 e^{-\Gamma_1 t} + |w_2|^2 e^{-\Gamma_2 t} \nonumber\\
&\quad + 2|w_1||w_2| \cos \left(\Delta_{12}t + \phi \right) e^{-(\Gamma_1+\Gamma_2)t/2},
\label{eq:pe_oscillation}
\end{align}
where $\Delta_{12} \equiv {\rm Re}\,\kappa_1 - {\rm Re}\,\kappa_2$ and $\phi$ is the argument of $w_1^\ast w_2$.
An oscillatory term appears as a cross-term.
The oscillation amplitude is characterized by the product of the weights, while the difference between the real parts of the eigenvalues determines the oscillation frequency.
Even when the weights are comparable, if the decay rates differ substantially, one of the two modes decays rapidly, and the oscillatory contribution is expected to be relevant only over a short period.
Hence, a problematic situation occurs when two (or more) eigenmodes have similarly large weights and similarly small decay rates.

Figure~\ref{fig:Gamma_Re}(b) shows two eigenmodes with large weights, whose decay rates are comparably small.
The difference in the real parts of the eigenvalues is $1.78$, indicating oscillations with period $T = 2\pi/1.78 \simeq 3.53$.
In Fig.~\ref{fig:Gamma_Re}(a), the dynamics exhibit approximately 14 oscillations up to $t=50$, being consistent with $50/T \simeq 14.16$.
Thus, oscillations in the excitation probability can be connected to the real parts of the eigenvalues.
The parameter selection in Fig.~\ref{fig:Gamma_Re} is adopted to illustrate a case with pronounced multimode oscillations.

Overall, regarding optimization, a desirable structure for local-excitation retention should exhibit a single mode with a large weight and small decay rate.

\subsection{Spectral surrogate for design and assessment}

The spectral observations presented above suggest a compact score for judging whether a structure is favorable for local-excitation retention. 
Therefore, we introduce a surrogate quantity based on modal decay rates and overlap weights.
This quantity can be used as both an interpretive descriptor for candidate geometries and objective in inverse design.
We consider its use as a design objective as an illustrative application.
To encode this design principle in a differentiable objective, we introduce the following cost function:
\begin{equation}
F = \alpha \sum_{\ell} p_{\ell} \log(\Gamma_{\ell}/\gamma_{0}) - \beta \sum_{\ell} p_{\ell}\log p_{\ell},
\label{eq:cost}
\end{equation}
where $p_{\ell} = |w_{\ell}| / \sum_{\ell'} |w_{\ell'}|$.

The first term evaluates the decay rates through their $p_\ell$-weighted logarithmic average.
This choice indicates that a small decay rate is beneficial only when the corresponding eigenmode substantially overlaps with the initial excitation.
In other words, reducing the decay rate of a mode with negligible weight does not notably improve the survival lifetime, whereas suppressing the decay of a mode with large weight directly lengthens the excitation survival.
Minimizing the first term favors configurations whose eigenmodes most relevant to the initial state are long-lived.
The logarithm is introduced because $\Gamma_\ell/\gamma_0$ typically spans several orders of magnitude, and a logarithmic scale provides a balanced comparison across subradiant and radiative modes.

The second term is the Shannon entropy of normalized weight distribution $\{p_\ell\}$.
Minimizing this entropy promotes the concentration of the weight in a small number of eigenmodes. In particular, it favors single-mode dominance, which is desirable because multimode weight sharing leads to interference-induced oscillations in $p_e(t)$.
We show $|w_\ell|^2$ because the diagonal contributions to the survival probability scale as $|w_\ell|^2 e^{-\Gamma_\ell t}$.
In the surrogate cost function, we use $|w_\ell|$ because the amplitude of the two-mode interference term is governed by products such as $|w_1||w_2|$, as described in Eq.~\eqref{eq:pe_oscillation}.

This cost function is neither unique nor intended as a rigorous bound on $p_e(t)$, but it establishes a physically motivated surrogate objective that captures the two abovementioned spectral requirements, namely, long lifetime and suppression of multimode interference.
In addition, Eq.~\eqref{eq:cost} compresses the phase-sensitive multimode dynamics into an amplitude-based spectral descriptor through normalized weights $\{p_\ell\}$. Thus, it should be interpreted as a ranking for candidate structures instead of a complete predictor of the full time dependence. In particular, we evaluate the final operational performance using time-domain quantity $p_e(t)$.
A quantitative comparison between the surrogate score and time-domain retention metric is presented in Appendix~A.

\subsection{Constrained inverse-design example}

\begin{figure*}[ht]
 \centering
 \includegraphics[width=0.9\linewidth]{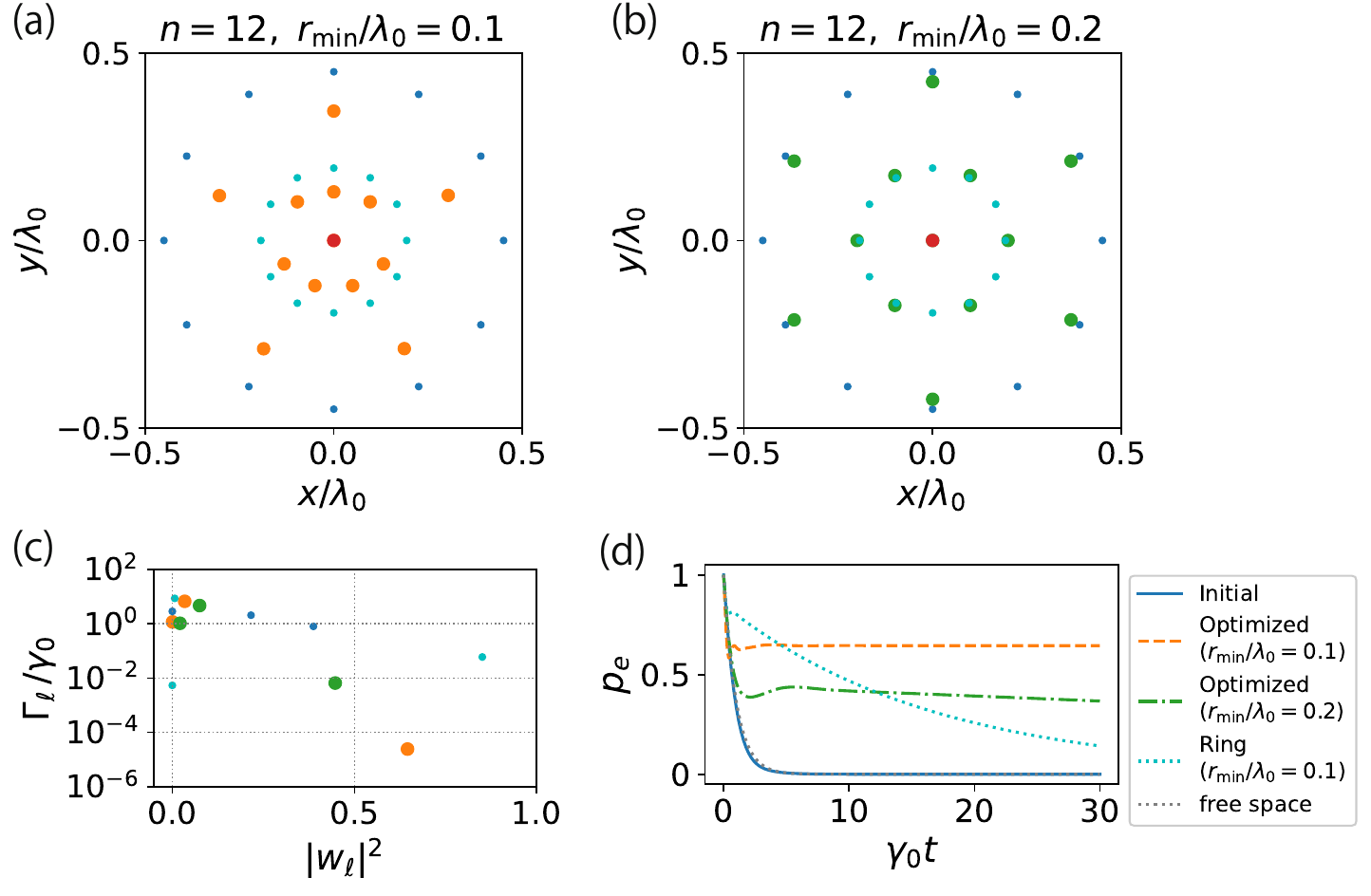}
 \caption{
 Optimized atomic configurations for $r_{\rm min}/\lambda_0$ of (a) $0.1$ and (b) $0.2$.
 The optimized structure depends on $r_{\rm min}$.
 For visual clarity, the optimized atomic configuration is shown after an $xy$-plane rotation that does not affect the dynamics.
 (c) Distribution of weights and decay rates. 
 The blue, orange, green, and cyan markers indicate the initial configuration, optimized structure for $r_{\rm min}/\lambda_0=0.1$, optimized structure for $r_{\rm min}/\lambda_0=0.2$, and reference ring configuration with $r_{\rm min}/\lambda_0=0.1$, respectively.
 For clarity, only the three markers with the largest weights are shown for each case.
 As expected, the weight after optimization is concentrated in eigenmodes with small decay rates.
 (d) Decay dynamics of storage atom showing improved local-excitation retention.
 The blue, orange, green, and cyan lines correspond to the same structures shown in panel (c).}
 \label{fig:ring_n12_best}
\end{figure*}

In practice, the achievable interatomic spacing is constrained by the choice of atomic species and lasers~\cite{Zhang2021, Ban2005}.
We impose minimum interatomic distance $r_{\rm min}$ as a nonlinear constraint in a constrained optimization scheme:
\begin{align}
|\br_{j} - \br_{j' \neq j}| \ge r_{\rm min}.
\end{align}

All atoms are assumed to share dipole orientation $\hat{\bd}=(1,i,0)/\sqrt{2}$.
As the initial configuration for optimization, we adopt a ring geometry (Fig.~\ref{fig:structure}), where 
an excited storage atom is placed at the center, and $n$ atoms in the ground state are arranged on a ring of radius $a\lambda_0$.
We consider $a=0.45$ and $n=12$ (see Appendix~B for results with $n=10$).
During optimization, the storage-atom position is fixed, and the design variables are the $xy$ coordinates of the $n$ surrounding atoms.
We perturb the surrounding atoms of the initial configuration by adding independent Gaussian noise with standard deviation $0.01\lambda_0$ to the $xy$ coordinates and then perform optimization.
From 100 runs with different perturbation patterns, we extract the structure with the minimum cost (see Appendix~C for the other optimization outputs).
For constrained optimization, we use the sequential least squares quadratic programming algorithm~\cite{Kraft1988} implemented in the Python \texttt{scipy} library.
For the minimum-distance constraint, we consider $r_{\rm min}/\lambda_0 \in \{0.1,\,0.2\}$.
Hyperparameters $\alpha$ and $\beta$ in Eq.~\eqref{eq:cost} adjust the tradeoff between decay-rate reduction and mode-weight concentration.
In this study, we set $(\alpha,\beta)=(1,3)$.
We use $p_e(t^\ast)$ as the final operational figure of merit, while the weight and decay rate are spectral descriptors explaining that performance. Time $t^\ast$ is defined by $\gamma_0 t^\ast = 30$.

Figures~\ref{fig:ring_n12_best}(a) and \ref{fig:ring_n12_best}(b) show the optimized atomic configurations for $r_{\rm min}/\lambda_0=0.1$ and $0.2$, respectively.
In both cases, optimization converges to a nontrivial configuration that differs from the initial ring geometry.
The optimized structures are different in both cases.
Hence, the optimized configuration depends on $r_{\rm min}$, which is an experimental constraint.
Our results suggest the effectiveness of the spectral criterion for leading to illustrative structures identified under experimentally motivated constraints.
As shown in Fig.~\ref{fig:ring_n12_best}(c), the optimized structures for $r_{\min}/\lambda_0=0.1$ and $0.2$ shift the dominant spectral weight toward eigenmodes with larger weights and smaller decay rates relative to the initial ring seed.
Hence, optimization concentrates the initial-state weight in longer-lived eigenmodes, being consistent with the design objective. Correspondingly, the excitation probability of the storage atom is enhanced, as shown in Fig.~\ref{fig:ring_n12_best}(d), indicating increased local-excitation retention within the present surrogate setting.

As listed in Table~\ref{tab:summary_n12}, a smaller $r_{\rm min}$ leads to a larger improvement in both the dominant weight and corresponding decay rate, with the largest weight moving further toward the lower-right region in Fig.~\ref{fig:ring_n12_best}(c).
The eigenmode decay rate reduces with respect to the largest weight by approximately two orders of magnitude.
Notably, a large dominant weight alone does not guarantee the best retention. For instance, the ring reference carries a large weight on one mode, but its associated decay rate is much larger than that in the optimized structures.
Therefore, within the considered parameter range, a smaller $r_{\rm min}$ promotes local-excitation retention.
We draw similar conclusions for other atom counts (see Appendix~B for results with $n=10$).
For a robustness analysis against positional fluctuations, see Appendix~D.

\begin{table}[ht]
  \centering
  \caption{Figures of merit of optimized and reference structures for $n=12$. Time $t^\ast$ is defined by $\gamma_0 t^\ast = 30$, and $\ell_{\max}$ denotes the index of the eigenmode with the largest weight.}
  \label{tab:summary_n12}
  \begin{tabular}{|c|c|c|c|c|} \hline
    Structure & $r_{\min} / \lambda_0$ & $|w_{\ell_{\max}}|^2$ & $\Gamma_{\ell_{\max}}/\gamma_0$ & $p_e(t^\ast)$ \\ \hline
    Optimized & $0.1$ & $0.65$ & $2.43 \times 10^{-5}$ & $0.65$ \\ \hline
    Optimized & $0.2$ & $0.45$ & $6.56 \times 10^{-3}$ & $0.37$ \\ \hline
    Square & $0.1$ & $0.002$ & $2.82 \times 10^{-3}$ & $0.01$ \\ \hline    
    Sunflower & $0.1$ & $0.23$ & $1.58 \times 10^{-3}$ & $0.21$ \\ \hline    
    Ring & $0.1$ & $0.85$ & $5.97 \times 10^{-2}$ & $0.14$ \\ \hline    
  \end{tabular}
\end{table}

\subsection{Modal and far-field interpretation of design example}

\begin{figure}[ht]
 \centering
 \includegraphics[width=0.99\linewidth]{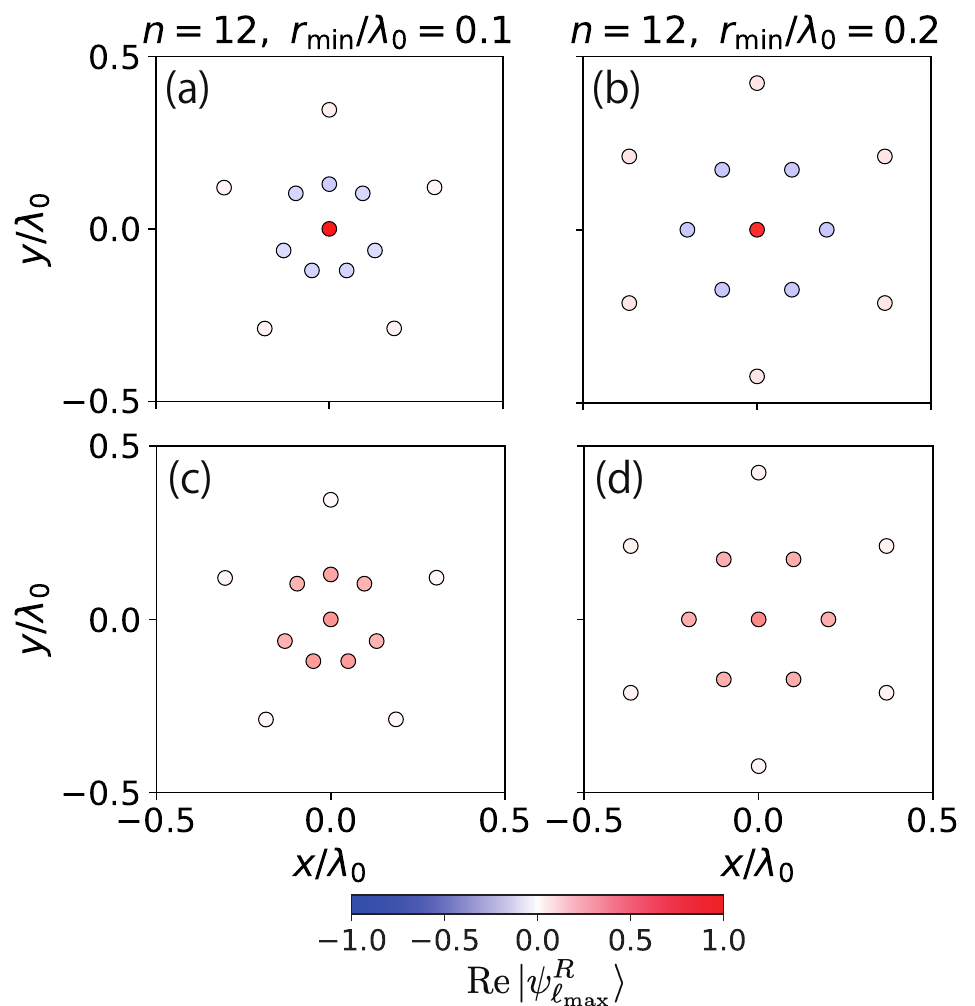}
 \caption{
 Real parts of eigenstates with the largest weights for $r_{\rm min}/\lambda_0$ of (a) $0.1$ and (b) $0.2$.
 The amplitude is strongly localized at the central storage atom.
 The distribution exhibits an alternating (out-of-phase) pattern, being positive on the central atom, negative on the inner surrounding atoms, and positive again on the outer atoms.
 This suggests that radiation in the far field is suppressed by destructive interference.
 (c), (d) Real parts of eigenstates with the maximum decay rates for $r_{\rm min}/\lambda_0$ of (c) $0.1$ and (d) $0.2$.
 Because the imaginary components are small ($< 10^{-2}$ for (a), (b), $< 10^{-1}$ for (c), and $< 0.14$ for (d)), they are omitted for clarity.} 
 \label{fig:eig_n12_best}
\end{figure}

To clarify the decay reduction by the optimized structures, we analyze the dominant eigenmodes and their far-field radiation.
Figures~\ref{fig:eig_n12_best}(a) and \ref{fig:eig_n12_best}(b) show the right eigenstate with the largest weight for $r_{\rm min}/\lambda_0=0.1$ and $0.2$, respectively.
Given its large weight, a large positive value is obtained at the central storage atom.
Moreover, the atoms are arranged in a concentric configuration, forming an out-of-phase pattern for the inner ring to have negative components, while the outer ring has positive components.
This suggests that destructive interference suppresses radiation in the far field.
Figures~\ref{fig:eig_n12_best}(c) and \ref{fig:eig_n12_best}(d) show the eigenstates with the maximum decay rate.
The phase-aligned components suggest constructive interference that enhances radiation in the far field.

In the absence of an incident input field, the positive-frequency component of the electric-field operator is given by
\begin{equation}
\hat{\bE}^{+}(\br) = \mu_0 \omega_0^2
\sum_{j=1}^{N} \bG_0(\br - \br_j, \omega_0) \cdot \bd \, \hat{\sigma}_{ge}^{j}.
\label{eq:Eplus_start}
\end{equation}
Using the far-field asymptotic form of the free-space Green's tensor 
for observation point $\mathbf r = r \hat{\br}$ sufficiently far from the atomic array,
we have
\begin{equation}
\bG_0(\br - \br_j, \omega_0)
\simeq \frac{e^{ik_0 |\br - \br_j|}}{4\pi |\br - \br_j|}
\left( \bI - \hat{\br} \hat{\br} \right),
\label{eq:G_far}
\end{equation}
where $k_0 = 2\pi / \lambda_0$ and $\bI$ denotes the identity matrix.
Furthermore, in far-field limit $r\gg |\br_j|$, we have
\begin{equation}
|\br - \br_j|
\simeq r - \hat{\br} \cdot \br_j,
\qquad \frac{e^{ik_0 |\br - \br_j|}}{|\br - \br_j|}
\simeq \frac{e^{ik_0 r}}{r}
e^{-ik_0 \hat{\br} \cdot \br_j}.
\label{eq:far_phase}
\end{equation}
Substituting this into Eq.~\eqref{eq:Eplus_start}, the far-field electric-field operator becomes
\begin{equation}
\hat{\mathbf E}^{+}_{\mathrm{ff}}(\br)
\simeq \frac{\mu_0 \omega_0^2}{4\pi} \frac{e^{ik_0 r}}{r}
\sum_{j=1}^{N} \left( \bI - \hat{\br} \hat{\br} \right) \cdot \bd \, \hat{\sigma}_{ge}^{j}
e^{-ik_0 \hat{\br} \cdot \br_j},
\label{eq:Eplus_ff_operator}
\end{equation}
where subscript $\mathrm{ff}$ denotes the far field.
Taking $\bra{\psi_\ell^R} \hat{\bE}^{-}_{\mathrm{ff}}(\br)
\hat{\bE}^{+}_{\mathrm{ff}}(\br) \ket{\psi_\ell^R}$ ($\hat{\bE}^{-}_{\mathrm{ff}}$ denotes the Hermitian conjugate of $\hat{\bE}^{+}_{\mathrm{ff}}$),
the far-field intensity of mode $\ell$ is obtained as
\begin{equation}
I_\ell(\hat{\br}) \propto
\left| \sum_{j=1}^{N} \left(
\bI - \hat{\br}\hat{\br} \right) \cdot \hat{\bd} \,
c_j^{(\ell)} e^{-ik_0 \hat{\br} \cdot \br_j} \right|^2
\equiv \bar{I}_\ell (\hat{\br}),
\label{eq:intensity_general}
\end{equation}
where $c_j^{(\ell)}$ is the complex amplitude of $\ell$-th eigenstate
$\ket{\psi_\ell^{R}} = \sum_{j=1}^{N} c_j^{(\ell)} \ket{e_j} \bigotimes_{j' \neq j} \ket{g_{j'}}$ and 
$\bar{I}_\ell (\hat{\br})$ denotes the dimensionless angular radiation pattern of mode $\ell$.
Equation~\eqref{eq:intensity_general} follows from the standard far-field radiation formula for an array of point dipoles,
with each atom contributing phase factor $e^{-i k_0 \hat{\br} \cdot \br_j}$ determined by its position.
We define the integrated dimensionless radiation pattern as
\begin{equation}
\bar{P}_\ell \equiv \int d\Omega\, \bar{I}_\ell(\hat{\br}),
\label{eq:integrated_pattern}
\end{equation}
where $d\Omega = \sin\theta\, d\theta\, d\phi$.
In numerical implementations, $\bar{P}_\ell$ is evaluated by a discrete quadrature on the sphere.
$\bar{I}_\ell$ and $\bar{P}_\ell$ characterize the dimensionless angular structure of radiation
rather than the absolute radiated power in physical units.
The integrated radiation pattern and radiative decay rate are proportional (see Appendix~E for the detailed derivation):
\begin{equation}
\frac{\Gamma_\ell}{\gamma_0} = \frac{3}{8\pi} \, \bar{P}_\ell.
\label{eq:gamma_P_relation}
\end{equation}
When comparing different eigenmodes, we fix the excitation amount by imposing
$\sum_{j=1}^{N}|c_j^{(\ell)}|^2=1$.
Thus, differences in $\bar{P}_\ell$ directly reflect differences in radiative coupling,
and the proportionality in Eq.~\eqref{eq:gamma_P_relation} holds without additional mode-dependent factors.

\begin{figure}[ht]
 \centering
 \includegraphics[width=0.85\linewidth]{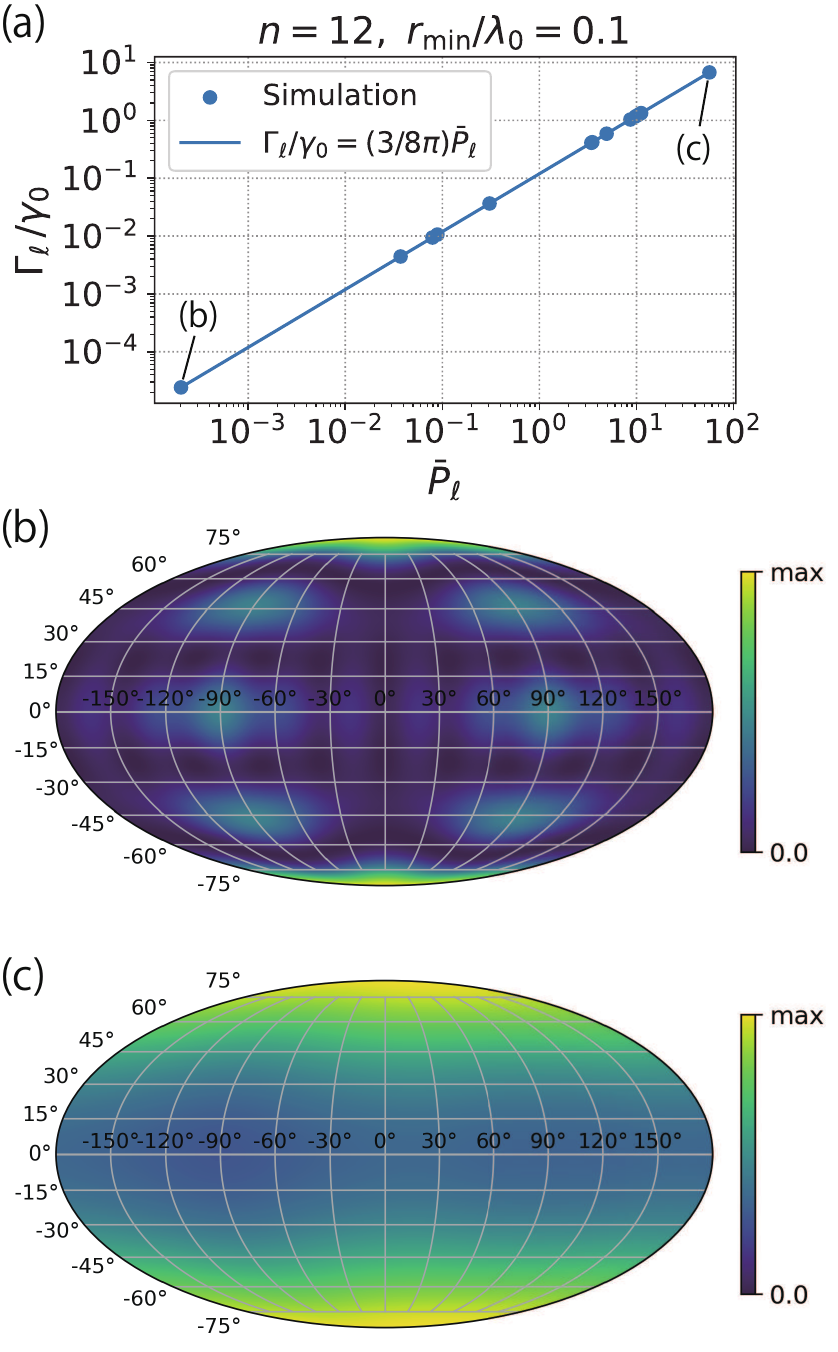}
 \caption{
 (a) Relationship between integrated dimensionless radiation pattern in far field and decay rate per eigenstate.
 A suitable agreement with the analytical expression is observed.
 Far-field radiation intensity of eigenstates with the (b) minimum and (c) maximum decay rates.
 The color indicates the intensity normalized by its maximum value.
 Compared with the eigenstate with the maximum decay rate, the eigenstate with the minimum decay rate exhibits suppressed far-field radiation over a wider angular range.
 }
 \label{fig:far-field_n12}
\end{figure}

Figure~\ref{fig:far-field_n12}(a) shows the relationship between the decay rate and integrated dimensionless radiation pattern $\bar{P}_\ell$ per eigenstate in the optimized structure for $r_{\rm min}=0.1$.
The numerical results are consistent with the analytical proportionality over the full set of eigenmodes.
The far-field radiation patterns of the eigenstates with the minimum and maximum decay rates are shown in Figs.~\ref{fig:far-field_n12}(b) and (c), respectively.
The mode with the maximum decay rate radiates over nearly the entire solid angle, mainly around the $z$ axis.
In contrast, omnidirectional radiation is mostly suppressed for the mode with the minimum decay rate.

The results indicate that the relevant design principle for local-excitation retention in impurity-assisted atomic arrays does not simply involve searching for a geometry with an extremely small minimum decay rate but rather to obtain a structure whose initial excitation is funneled predominantly into a single long-lived collective mode while suppressing competing modes that induce oscillatory dynamics.
Accordingly, minimum-distance constraint $r_{\rm min}$ plays a dual role by limiting the achievable degree of subradiance while selecting the class of experimentally constrained geometries. 
Therefore, the results suggest that spectral quantities (i.e., decay rates and weights) provide a compact yet physically transparent set of descriptors for the inverse design of impurity-assisted structures with enhanced local-excitation retention.
This perspective also motivates the broader use of spectrum-based optimization in cooperative light--matter systems that require simultaneous balance of a long lifetime, robustness against multimode interference, and experimental feasibility.

\section{Conclusion}
\label{sec:Conclusion}

We investigate the survival of an initially localized excitation in an impurity-assisted atomic array through the eigenmode decomposition of an effective non-Hermitian Hamiltonian.
By examining representative geometries, we identify spectral mechanisms that control the dynamics.
Remarkably, the smallest collective decay rate is not sufficient to characterize the performance. Instead, the relevant factor is the combination of the decay rates and overlaps of the corresponding eigenmodes with the initial excitation.
In addition, pronounced oscillations occur when two or more long-lived modes carry comparable weights, with oscillation frequencies set by differences in the real parts of the eigenvalues.

These observations unveil a practical design principle for impurity-assisted local-excitation retention. In detail, the initial excitation should be funneled predominantly into a single long-lived collective mode, while competing modes responsible for multimode interference should be suppressed. 
The introduced surrogate provides a compact way to summarize these competing spectral requirements. As a proof of principle, we embed this surrogate for constrained atom-position optimization that provides aperiodic configurations with improved retention.
The optimized structures depend on the allowed minimum spacing, and smaller $r_{\rm min}$ leads to stronger subradiance and improved performance within the considered parameter range.

Our performance measure is the survival probability of the initially prepared one-excitation state rather than the efficiency or fidelity of a complete write--store--read protocol, and the spectral objective establishes a physically motivated surrogate rather than a rigorous bound on the dynamics.
While we address robustness against positional fluctuations, other experimentally relevant imperfections remain to be explored.
Important next steps include disorder in dipole alignment and atomic detuning, larger system sizes, and the incorporation of explicit write and retrieval processes.
With these extensions, the present spectral viewpoint may provide a useful basis toward photon-storage or quantum-memory protocols~\cite{Garcia2017, Rastogi2022, Bonanno2022, Facchinetti2016, Ballantine2021, Rubies-Bigorda2022}.

\begin{acknowledgments}
J.O. thanks Takahiro Horiba for fruitful discussions on atomic configurations, Yuki Sato for helpful comments on numerical optimization techniques, and Seiji Kajita for constructive comments that helped improve the manuscript.
\end{acknowledgments}

\appendix
\section{Quantitative assessment of spectral surrogate}

To examine the description by the surrogate objective in Eq.~\eqref{eq:cost} of time-domain retention, we generated 100 candidate structures from the ring configuration for $r_{\rm min}/\lambda_0=0.1$ by adding positional fluctuations to the $xyz$ coordinates.
The amplitude of the fluctuation was $0.01\lambda_0$.
Then, we evaluated both surrogate value $F$ and operational metrics $p_e(t^\ast)$ and $\bar p_e(t^\ast)$.
Time-averaged excitation probability $\bar{p}_e(t^\ast)$ is defined as
\begin{equation}
\bar{p}_e (t^\ast)=\frac{1}{t^\ast}\int_{0}^{t^\ast} p_e(t)\,dt.
\end{equation}

\begin{figure}[ht]
 \centering
 \includegraphics[width=0.9\linewidth]{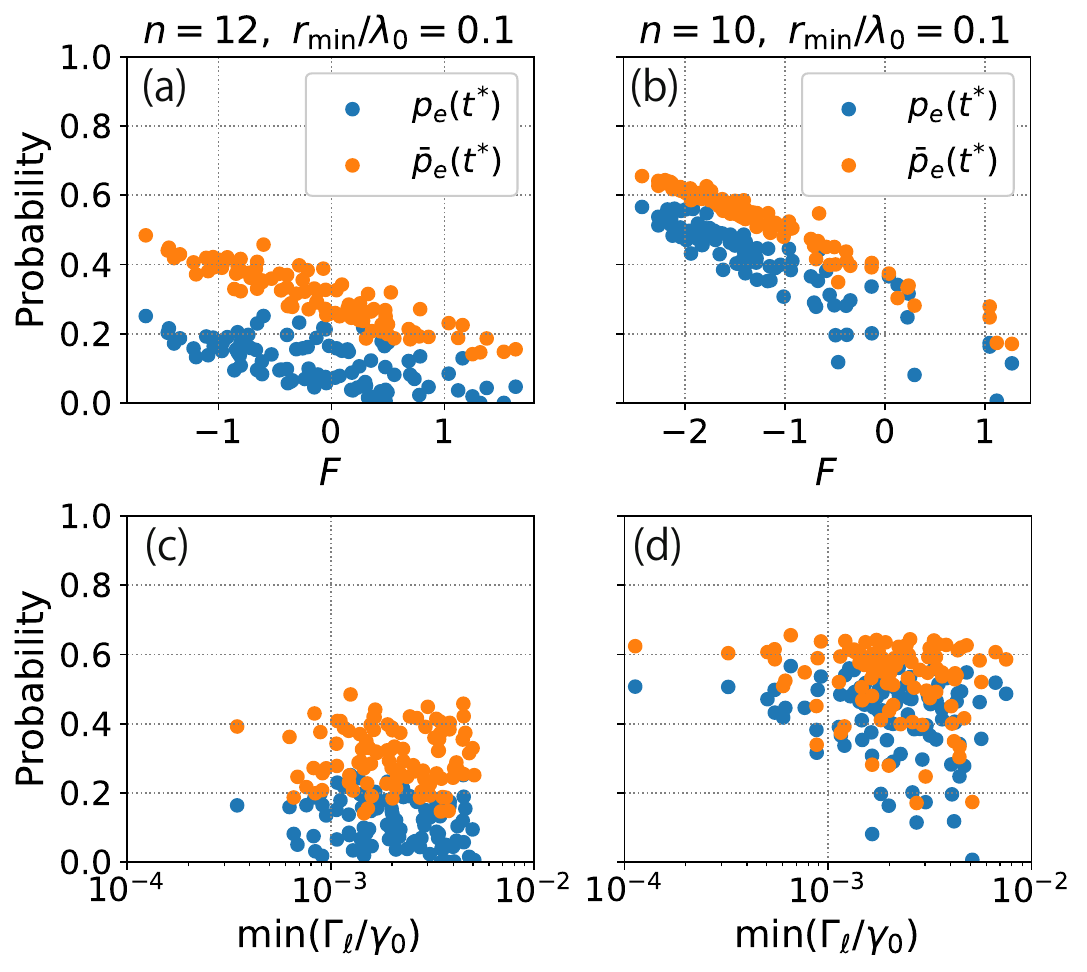}
 \caption{
 Scatter plots comparing surrogate score $F$ with time-domain retention metrics $\{p_e(t^\ast)$ and $\bar{p}_e(t^\ast)\}$ for 100 candidate structures and $n$ of (a) $12$ and (b) $10$ surrounding atoms.
 Each point represents one structure.
 Structures with lower $F$ tend to exhibit higher time-domain retention.
 Panels (c) and (d) are the same as panels (a) and (b) but for objective $\min(\Gamma_\ell / \gamma_0)$.}
 \label{fig:surrogate}
\end{figure}

Figures~\ref{fig:surrogate}(a) and \ref{fig:surrogate}(b) show the relationship between the surrogate value and operational metrics.
Although Eq.~\eqref{eq:cost} is not a rigorous bound and does not retain the full phase information contained in complex weights $w_\ell$, lower $F$ is correlated with higher time-domain performance.
In contrast, Figs.~\ref{fig:surrogate}(c) and \ref{fig:surrogate}(d) show that the minimum decay rate, $\min(\Gamma_\ell/\gamma_0)$, alone is not correlated with performance.

Table~\ref{tab:corr} shows that surrogate objective $F$ is strongly correlated with the two time-domain retention metrics, whereas the minimum decay rate, $\min(\Gamma_\ell/\gamma_0)$, shows a weak correlation.
In particular, the Pearson correlation coefficients between $F$ and either $p_e(t^\ast)$ or $\bar{p}_e(t^\ast)$ are consistently large, while those for $\min(\Gamma_\ell/\gamma_0)$ are small.
Hence, $F$ is a promising practical ranking criterion for candidate structures.
The correlation is systematically stronger for time-averaged metric $\bar{p}_e(t^\ast)$ than for single-time metric $p_e(t^\ast)$.
Because $p_e(t)$ can exhibit oscillations owing to multimode interference, its value at final time $t^\ast$ may show additional scatter depending on the oscillation phase. In contrast, average $\bar{p}_e(t^\ast)$ is less sensitive to such phase-dependent fluctuations, robustly reflecting the spectral quality of a structure. 

\begin{table}[ht]
\centering
\caption{Correlation coefficients between surrogate value and operational metrics.}
\label{tab:corr}
\begin{tabular}{|c|c|c|c|}
\hline
Structure & Objective & Metric & Pearson correlation \\
\hline
Ring ($n=12$) & $F$ & $p_e(t^\ast)$        & $-$0.59 \\ \cline{3-4}
                        &        & $\bar p_e(t^\ast)$ & $-$0.91 \\ \cline{2-4}
                        & $\min(\Gamma_\ell/\gamma_0)$ & $p_e(t^\ast)$ & $-$0.22 \\ \cline{3-4}
                        &        & $\bar p_e(t^\ast)$ & 0.02 \\ \hline                        
Ring ($n=10$) & $F$ & $p_e(t^\ast)$        & $-$0.90 \\ \cline{3-4}
                        &        & $\bar p_e(t^\ast)$ & $-$0.98 \\ \cline{2-4}
                        & $\min(\Gamma_\ell/\gamma_0)$ & $p_e(t^\ast)$ & $-$0.18 \\ \cline{3-4}
                        &        & $\bar p_e(t^\ast)$ & $-$0.16 \\                        
\hline
\end{tabular}
\end{table}

\section{Results for $n=10$ surrounding atoms}

\begin{figure*}[ht]
 \centering
 \includegraphics[width=0.6\linewidth]{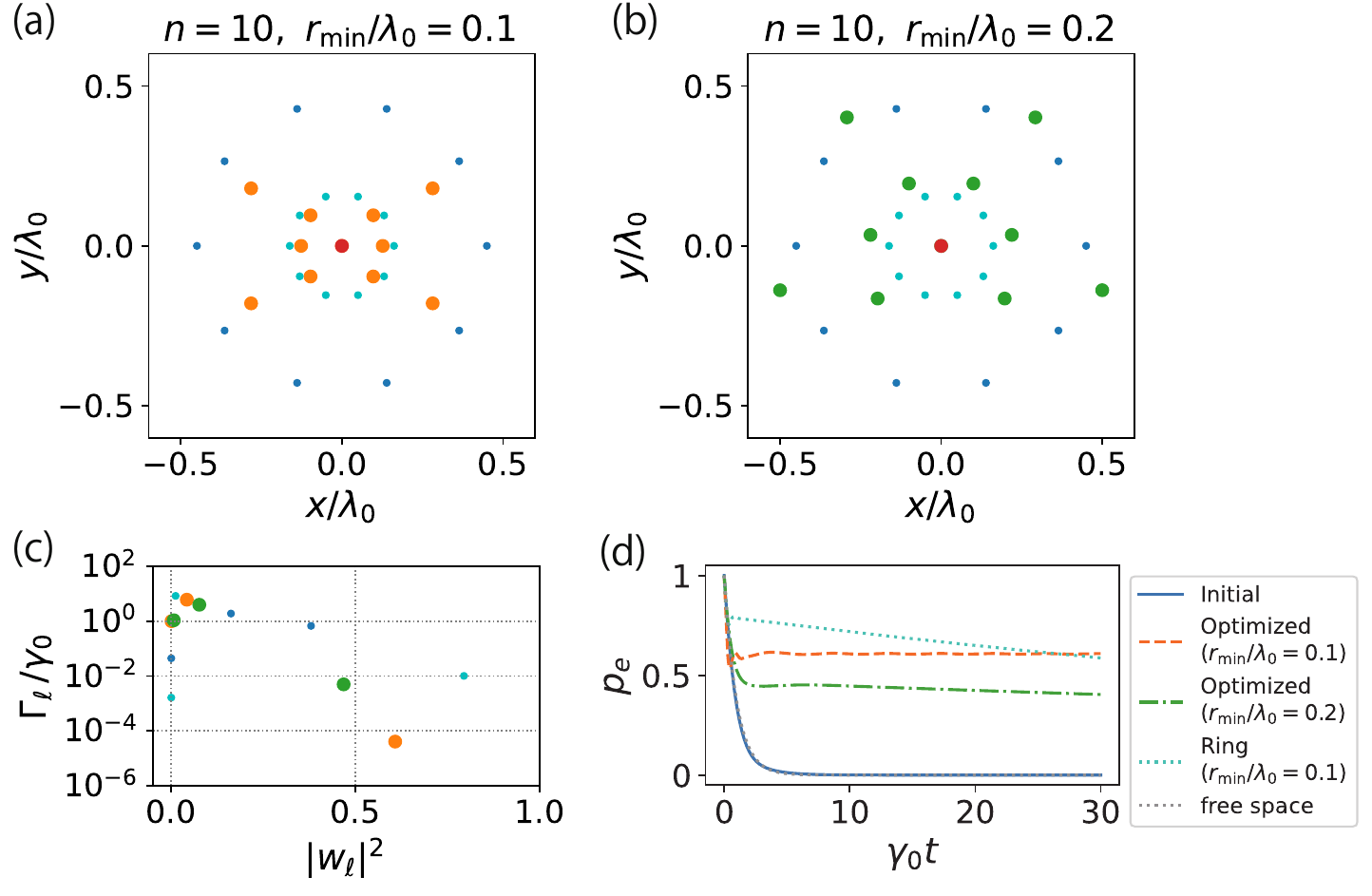}
 \caption{
 Same as Fig.~\ref{fig:ring_n12_best} but for $n=10$ atoms.}
 \label{fig:ring_n10_best}
\end{figure*}

Figure~\ref{fig:ring_n10_best} is analogous to Fig.~\ref{fig:ring_n12_best} but for $10$ surrounding atoms ($n=10$).
Table~\ref{tab:summary_n10} lists the figures of merit for the optimized structures.
Like for $n=12$, optimization converges to different structures depending on the minimum interatomic distance.
Comparing the excitation probability at the final time, $p_e(t^\ast)$, $n=12$ surrounding atoms yield higher values than $n=10$.
The degree of improvement depends strongly on $n$ and on the spacing constraint.
For $n=10$ and $r_{\rm min}/\lambda_0=0.1$, the optimized structure shows a marginal improvement in $p_e(t^\ast)$ over the ring reference.

\begin{table}[ht]
  \centering
  \caption{Figures of merit of optimized and reference structures for $n=10$. Time $t^\ast$ is defined by $\gamma_0 t^\ast = 30$, and $\ell_{\max}$ denotes the index of the eigenmode with the largest weight.}
  \label{tab:summary_n10}
  \begin{tabular}{|c|c|c|c|c|} \hline
    Structure & $r_{\min} / \lambda_0$ & $|w_{\ell_{\max}}|^2$ & $\Gamma_{\ell_{\max}}/\gamma_0$ & $p_e(t^\ast)$ \\ \hline
    Optimized & $0.1$ & $0.61$ & $4.07 \times 10^{-5}$ & $0.61$ \\ \hline
    Optimized & $0.2$ & $0.47$ & $4.96 \times 10^{-3}$ & $0.40$ \\ \hline
    Square & $0.1$ & $0.004$ & $4.61 \times 10^{-3}$ & $0.02$ \\ \hline    
    Sunflower & $0.1$ & $0.23$ & $1.56 \times 10^{-3}$ & $0.22$ \\ \hline    
    Ring & $0.1$ & $0.79$ & $1.01 \times 10^{-2}$ & $0.59$ \\ \hline    
  \end{tabular}
\end{table}

\begin{figure}[ht]
 \centering
 \includegraphics[width=0.9\linewidth]{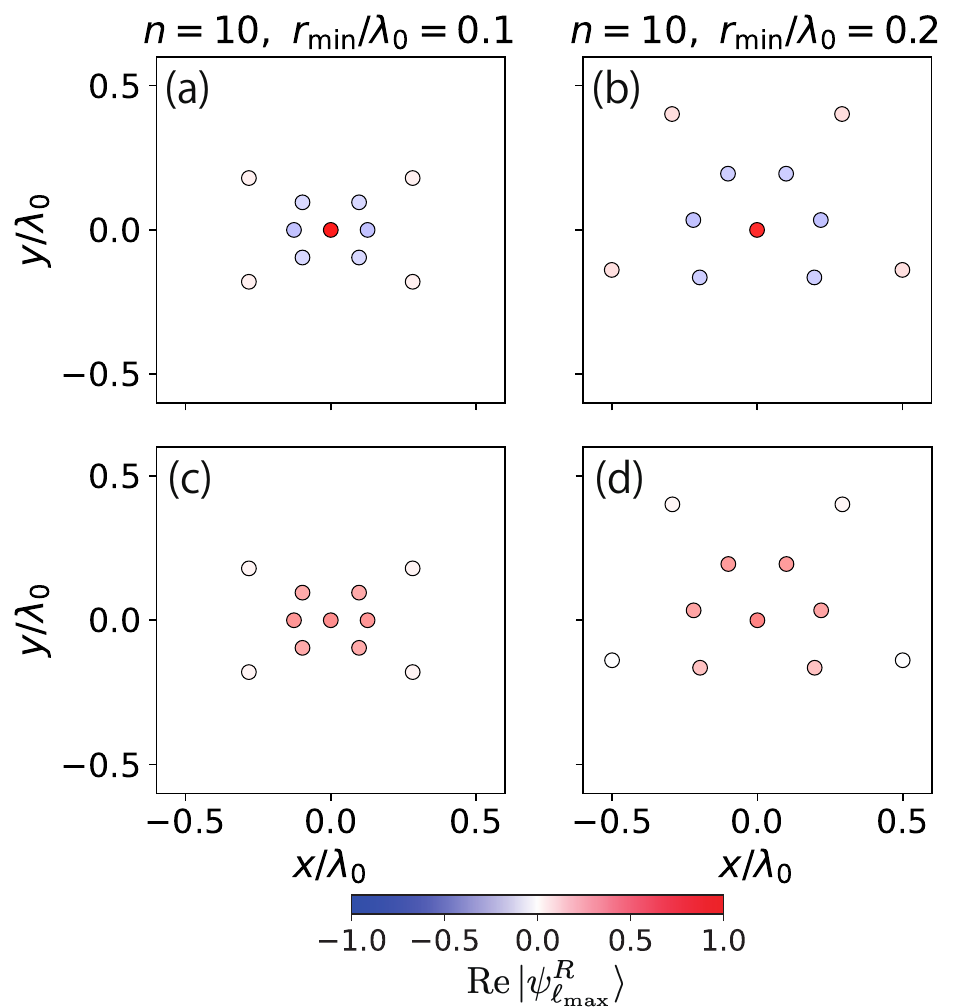}
 \caption{
  Same as Fig.~\ref{fig:eig_n12_best} but for $n=10$ atoms.}
 \label{fig:eig_n10_best}
\end{figure}

\begin{figure}[ht]
 \centering
 \includegraphics[width=0.8\linewidth]{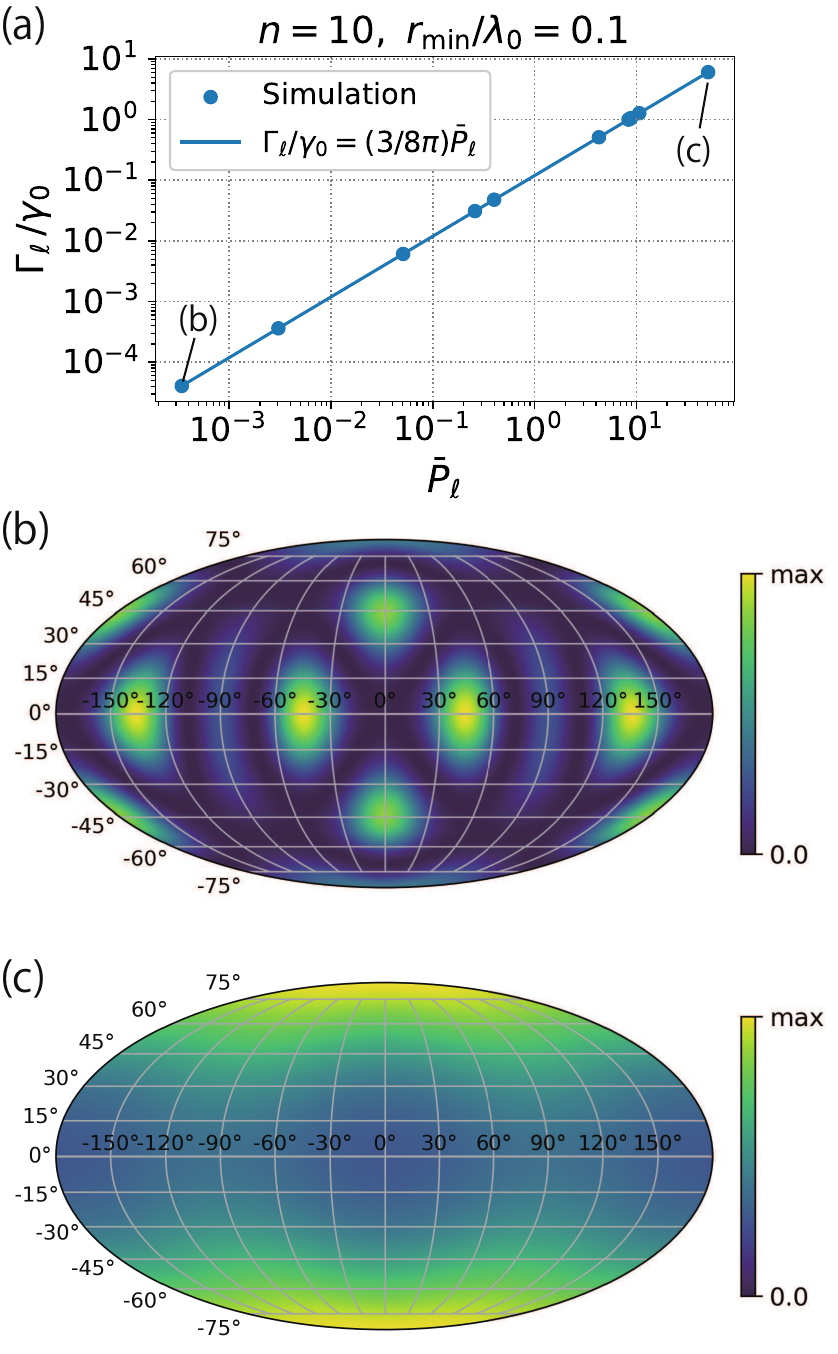}
 \caption{
 Same as Fig.~\ref{fig:far-field_n12} but for $n=10$.
 }
 \label{fig:far-field_n10}
\end{figure}

Figures~\ref{fig:eig_n10_best} and \ref{fig:far-field_n10} are analogous to Figs.~\ref{fig:eig_n12_best} and \ref{fig:far-field_n12} and support the same physical interpretation for the $n=10$ case.
The right eigenstates with the largest weights in Figs.~\ref{fig:eig_n10_best}(a) and~\ref{fig:eig_n10_best}(b) exhibit an out-of-phase pattern centered on the storage atom, indicating that destructive interference suppresses radiation in the far field.
Consistently, Fig.~\ref{fig:far-field_n10} shows that the eigenstate with the minimum decay rate exhibits more strongly suppressed far-field radiation over a wider angular range than the eigenstate with the maximum decay rate.

\section{Dependence on perturbed ring seeds}

\begin{figure}[ht]
 \centering
 \includegraphics[width=0.99\linewidth]{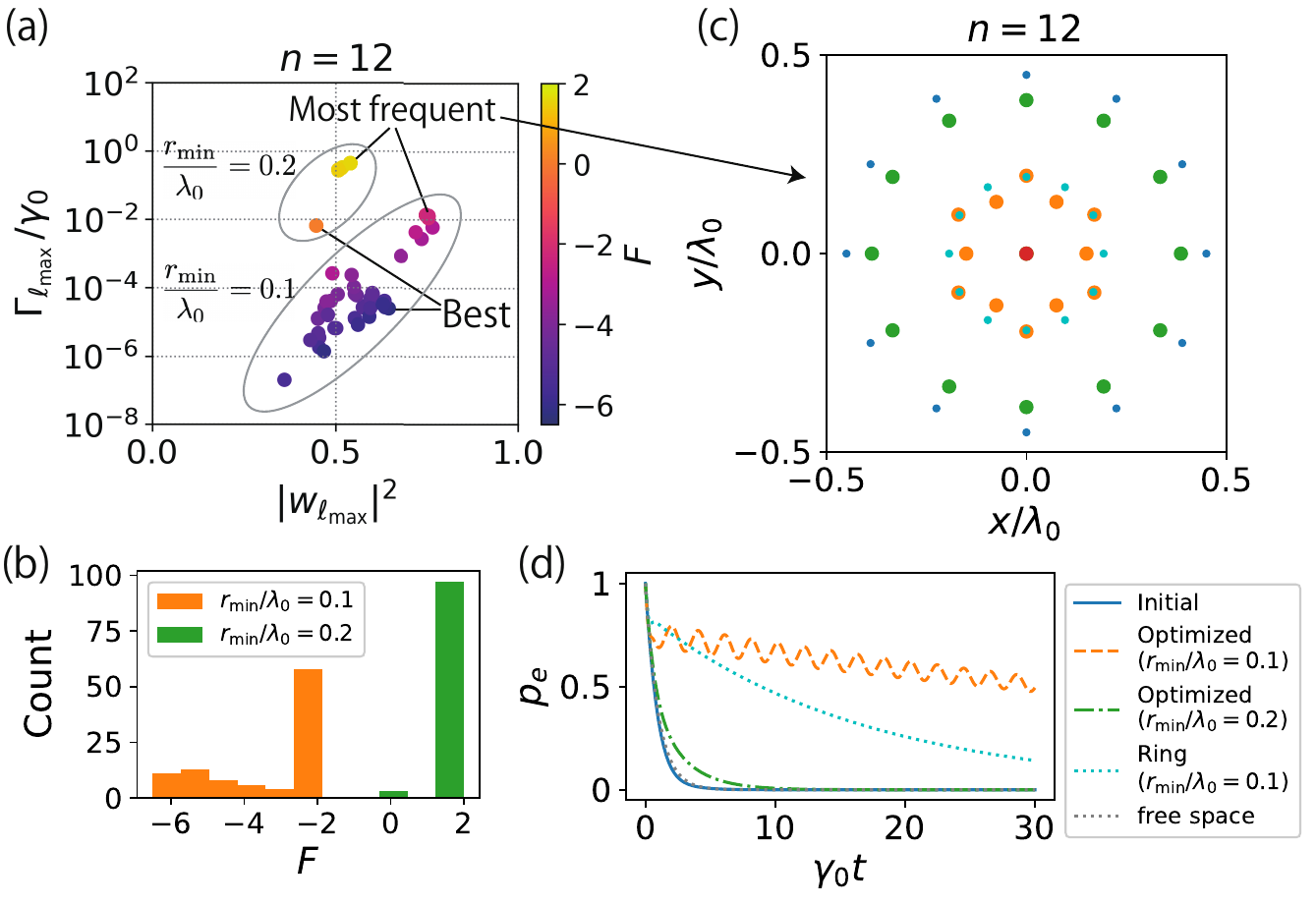}
 \caption{
 (a) Distribution of weight and decay rate of eigenmode with the largest weight in optimized structures.
 The results are obtained from 100 runs starting from different perturbed ring seeds under $r_{\rm min}/\lambda_0=0.1$ and $0.2$.
 The colormap encodes the cost-function value.
 (b) Distribution of cost-function values.
 (c) Most frequent structure and (d) its decay dynamics.
 Blue dots/lines indicate the initial configuration, and orange dots/lines indicate the optimized result.}
 \label{fig:ring_n12_mode}
\end{figure}

To examine the dependence of the optimization outcome on the initial seed, Fig.~\ref{fig:ring_n12_mode}(a) shows the distribution of the largest weight, $w_{\ell_{\rm max}}$, and related decay rate $\Gamma_{\ell_{\rm max}}$ obtained from 100 optimization runs with different initial configurations.
Figure~\ref{fig:ring_n12_mode}(b) shows the distribution of the cost-function values.
Most trials converge to a similar value, likely owing to constraint $r_{\rm min}/\lambda_0=0.2$ that drastically reduces the size of the search space.
The most frequent structure is a configuration with a smaller effective ring radius, as shown in Fig.~\ref{fig:ring_n12_mode}(c) (label ``Most frequent'').
Figure~\ref{fig:ring_n12_mode}(d) shows that the improvement in local-excitation retention is limited for this structure.

For the smaller value of $r_{\rm min}/\lambda_0=0.1$, because the constraint is relaxed and the search space broadens, the cost-function values span a wider range and reach smaller values than those for $r_{\rm min}/\lambda_0=0.2$ (Fig.~\ref{fig:ring_n12_mode}(a)).
This multiobjective optimization balancing decay-rate minimization and weight maximization is expected to have widely spread solutions toward the lower-right region as Pareto-optimal points in Fig.~\ref{fig:ring_n12_mode}(a).
The decay dynamics for the most frequent structure shown in Fig.~\ref{fig:ring_n12_mode}(c) exhibit small oscillations, while the excitation probability remains as high as approximately $50\%$, even at the final time (see Fig.~\ref{fig:ring_n12_mode}(d)).
Because this structure has a relatively high symmetry, it may offer practical advantages for experimental implementation, including simpler design and alignment.

\begin{figure}[ht]
 \centering
 \includegraphics[width=0.99\linewidth]{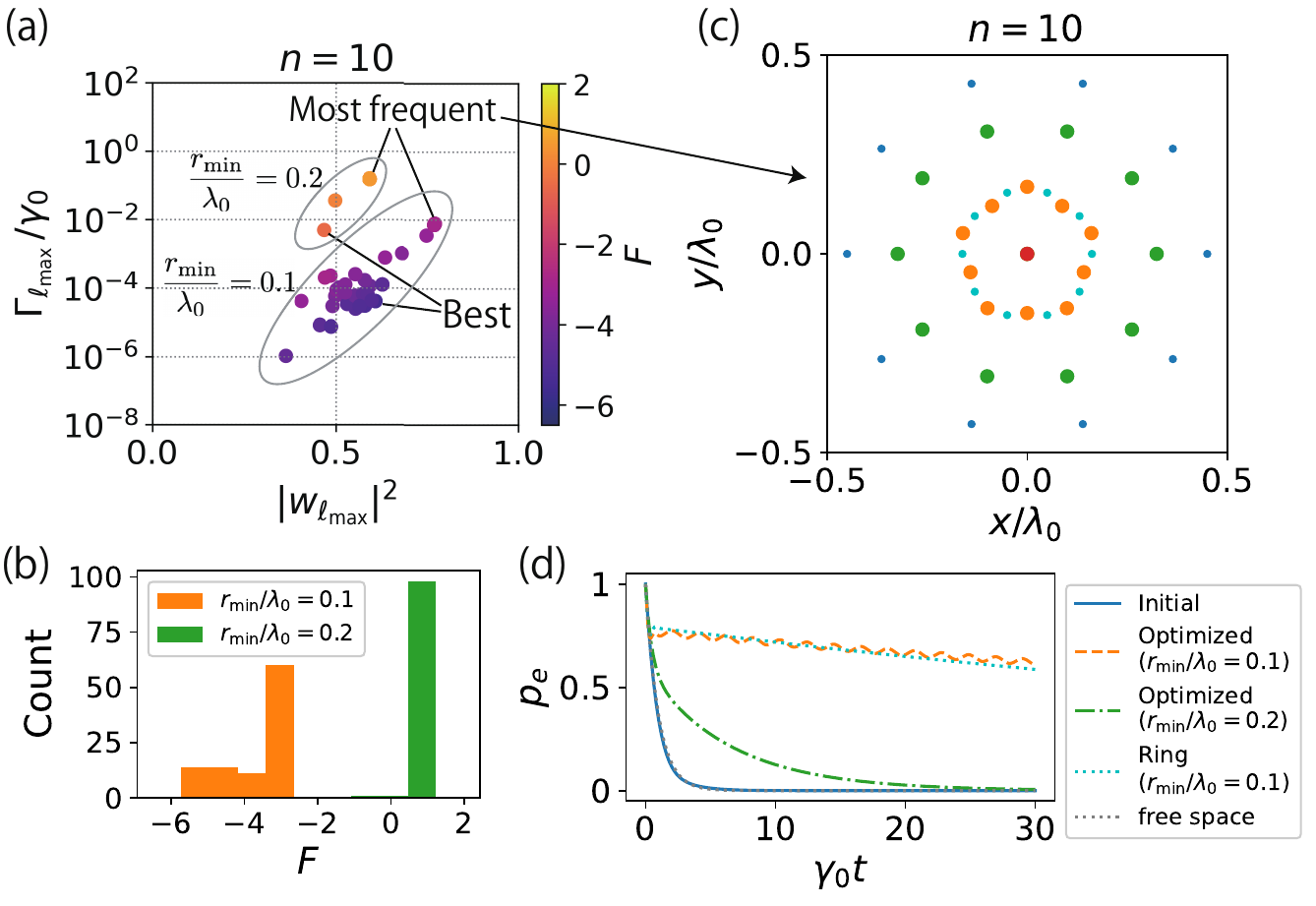}
 \caption{
 Same as Fig.~\ref{fig:ring_n12_mode} but for $n=10$.}
 \label{fig:ring_n10_mode}
\end{figure}

Figure~\ref{fig:ring_n10_mode} shows the corresponding dependence on the perturbed ring seeds for $n = 10$.
As in the $n = 12$ case, the results for $r_{\min}/\lambda_0 = 0.2$ are concentrated in a relatively narrow range, whereas those for $r_{\min}/\lambda_0 = 0.1$ are more broadly distributed and reach lower cost-function values, reflecting the enlarged search space under the relaxed minimum-distance constraint.
The most frequent structure in Fig.~\ref{fig:ring_n10_mode}(c) remains relatively symmetric, and its decay dynamics in Fig.~\ref{fig:ring_n10_mode}(d) indicate that the qualitative seed dependence is similar to that for $n = 12$.

\section{Robustness against positional fluctuations}

\begin{figure*}[ht]
 \centering
 \includegraphics[width=0.9\linewidth]{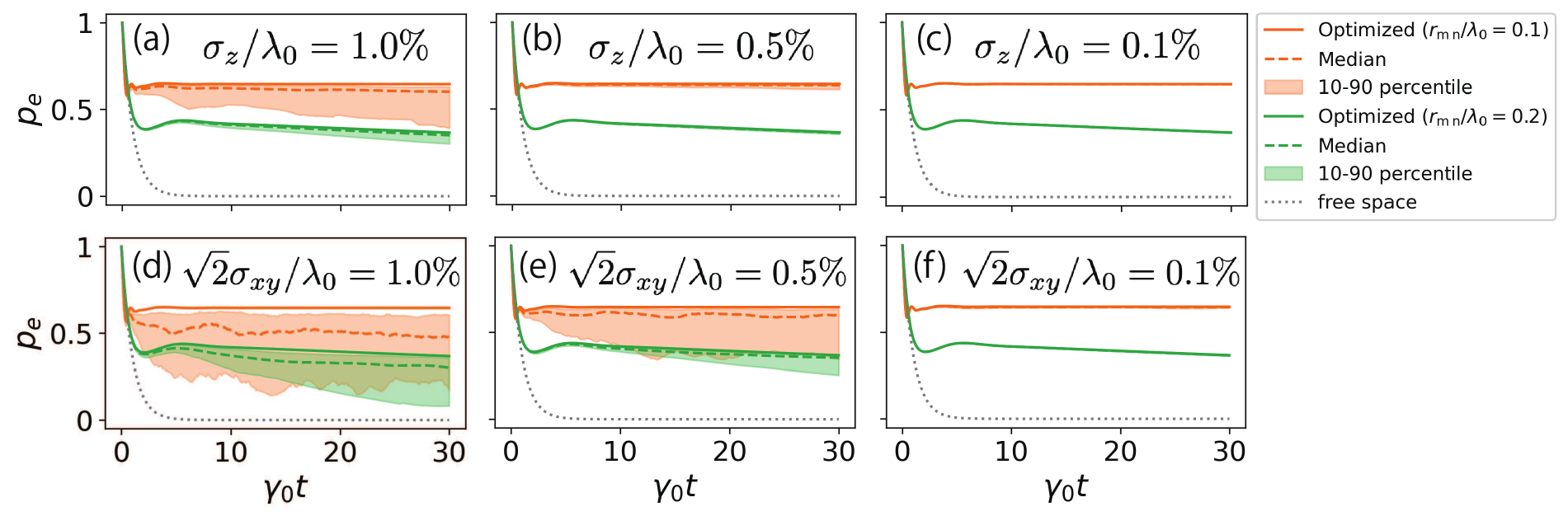}
 \caption{
 Excitation dynamics of optimized structures (solid lines).
 The dashed lines show the median over 100 fluctuation realizations, and the shaded regions indicate the 10--90 percentile range.
 Orange and green correspond to $r_{\rm min}/\lambda_0=0.1$ and $0.2$, respectively.}
 \label{fig:ring_n12_robust}
\end{figure*}

To perform a more realistic analysis under experimental conditions, we examine the robustness against atomic positional fluctuations.
Specifically, we add random fluctuations with amplitudes of 1\%, 0.5\%, and 0.1\% of the typical wavelength scale to the in-plane ($x$, $y$) and out-of-plane ($z$) coordinates of each atom and compare the resulting decay dynamics.
For the out-of-plane case, we add Gaussian fluctuations only to the $z$ coordinate with standard deviation $\sigma_z$.
For the in-plane case, Gaussian fluctuations are added to the $x$ and $y$ coordinates.
To ensure a fair comparison in the total fluctuation magnitude,
we set the per-axis standard deviation to $\sigma_{xy} = \sigma_z / \sqrt{2}$,
such that $\langle \delta x^2 + \delta y^2 \rangle = \langle \delta z^2 \rangle$.
For each fluctuation amplitude, 100 independent realizations are evaluated, and the mean and standard deviation of the excitation probability of the storage atom are calculated.
The results are shown in Fig.~\ref{fig:ring_n12_robust}.
Regardless of the fluctuation amplitude, the mean remains higher for the optimized structure. Moreover, for 0.1\% fluctuation, the deviation from the optimized value remains within 1\%.

\section{Relation between $\Gamma_\ell$ and $P_\ell$}

Let the right eigenstate of the effective non-Hermitian Hamiltonian be
\begin{equation}
\ket{\psi_\ell^{R}}
= \sum_{j=1}^{N} c_j^{(\ell)} \ket{e_j} \bigotimes_{j'\neq j} \ket{g_{j'}},
\end{equation}
and fix the normalization as
\begin{equation}
\sum_{j=1}^{N} |c_j^{(\ell)}|^2 = 1.
\end{equation}
Hence, the collective decay rate of the mode is given by the expectation value of the dissipative part:
\begin{equation}
\Gamma_\ell = \sum_{j,j'=1}^{N}
c_j^{(\ell)\,*}\, \Gamma_{jj'}\, c_{j'}^{(\ell)} .
\label{eq:D1}
\end{equation}
Using Eq.~\eqref{eq:Gamma} and assuming identical atoms with the same dipole orientation, $\hat{\bd}$, we obtain
\begin{equation}
\frac{\Gamma_\ell}{\gamma_0} = \frac{6\pi}{\omega_0}
\sum_{j,j'=1}^{N} c_j^{(\ell)\,*}
\left[ \hat{\bd}^{*} \cdot {\rm Im}\,\bG_0(\br_j - \br_{j'}, \omega_0) \cdot \hat{\bd} \right] c_{j'}^{(\ell)}.
\label{eq:D2}
\end{equation}

Next, we use the angular-spectrum representation of the imaginary part of the vacuum Green's tensor:
\begin{equation}
{\rm Im}\,\bG_0(\br, \omega_0) = \frac{\omega_0}{16\pi^2}
\int d\Omega\, \left(\bI - \hat{\br}\hat{\br}\right)
e^{i k_0 \hat{\br}\cdot \br},
\label{eq:D3}
\end{equation}
where $k_0=2\pi/\lambda_0$. Substituting Eq.~\eqref{eq:D3} into Eq.~\eqref{eq:D2} yields
\begin{align}
\frac{\Gamma_\ell}{\gamma_0}
&= \frac{3}{8\pi} \int d\Omega
\sum_{j,j'=1}^{N} c_j^{(\ell)\,*} c_{j'}^{(\ell)} \,
\hat{\bd}^{*}\cdot \left(\bI - \hat{\br}\hat{\br}\right)
\cdot \hat{\bd}\, e^{i k_0 \hat{\br}\cdot(\br_j - \br_{j'})} \nm\\
&= \frac{3}{8\pi} \int d\Omega\,
\left| \sum_{j=1}^{N} \left(\bI - \hat{\br}\hat{\br}\right) \hat{\bd}\, c_j^{(\ell)}
e^{-i k_0 \hat{\br}\cdot \br_j} \right|^2.
\label{eq:D4}
\end{align}
In the second line, as $\bI - \hat{\br}\hat{\br}$ is a Hermitian projection operator,
the double sum can be written as the squared modulus of the far-field amplitude.

On the other hand, the dimensionless angular radiation pattern introduced in Eq.~\eqref{eq:intensity_general} is given by
\begin{equation}
\bar{I}_\ell(\hat{\br}) = \left| \sum_{j=1}^{N}
\left(\bI - \hat{\br}\hat{\br}\right)\hat{\bd}\, c_j^{(\ell)}
e^{-i k_0 \hat{\br}\cdot \br_j} \right|^2,
\end{equation}
and its angular integral is defined by Eq.~\eqref{eq:integrated_pattern}:
\begin{equation}
\bar{P}_\ell = \int d\Omega\, \bar{I}_\ell(\hat{\br}).
\end{equation}
Therefore, Eq.~\eqref{eq:D4} becomes
\begin{equation}
\frac{\Gamma_\ell}{\gamma_0} = \frac{3}{8\pi} \bar{P}_\ell ,
\end{equation}
which is Eq.~\eqref{eq:gamma_P_relation}.


\begin{thebibliography}{99}

\bibitem{Novotny2012}
L. Novotny, B. Hecht, Principles of Nano-Optics. 2nd ed., Cambridge University Press (2012).

\bibitem{Chang2018}
D. E. Chang et al., Colloquium: Quantum matter built from nanoscopic lattices of atoms and photons,
Rev. Mod. Phys. {\bf 90}, 031002 (2018).


\bibitem{Dicke1954}
R. H. Dicke, Coherence in Spontaneous Radiation Processes, Phys. Rev. {\bf 93}, 99 (1954).

\bibitem{Lehmberg1970}
R. H. Lehmberg, Radiation from an $N$-Atom System. I. General Formalism, Phys. Rev. A {\bf 2}, 883 (1970).

\bibitem{Jen2025}
H. H. Jen, Photon-mediated dipole-dipole interactions as a resource for quantum science and technology in cold atoms, Quantum Sci. Technol. {\bf 10}, 023001 (2025).


\bibitem{Gross1982}
M. Gross and S. Haroche, Superradiance: An essay on the theory of collective spontaneous emission, Phys. Rep. {\bf 93}, 301 (1982).

\bibitem{Masson2022}
S. J. Masson and A. Asenjo-Garcia, Universality of Dicke superradiance in arrays of quantum emitters, Nat. Commun. {\bf 13}, 2285 (2022).

\bibitem{Sierra2022}
E. Sierra, S. J. Masson, and A. Asenjo-Garcia, Dicke superradiance in ordered lattices: Dimensionality matters, Phys. Rev. Res. {\bf 4}, 023207 (2022).

\bibitem{Rubies-Bigorda2023a}
O. Rubies-Bigorda, S. Ostermann, and S. F. Yelin, Characterizing superradiant dynamics in atomic arrays via a cumulant expansion approach, Phys. Rev. Res. {\bf 5}, 013091 (2023).


\bibitem{Bienaime2012}
T. Bienaim\'{e}, N. Piovella, and R. Kaiser, Controlled Dicke subradiance from a large cloud of two-level systems, Phys. Rev. Lett. {\bf 108}, 123602 (2012).

\bibitem{Garcia2017}
A. Asenjo-Garcia et al., Exponential Improvement in Photon Storage Fidelities Using Subradiance and ``Selective Radiance" in Atomic Arrays, Phys. Rev. X {\bf 7}, 031024 (2017).

\bibitem{Moreno-Cardoner2019}
M. Moreno-Cardoner, D. Plankensteiner, L. Ostermann, D. E. Chang, and H. Ritsch, Subradiance-enhanced excitation transfer between dipole-coupled nanorings of quantum emitters, Phys. Rev. A {\bf 100}, 023806 (2019).

\bibitem{Rubies-Bigorda2023b}
O. Rubies-Bigorda, S. Ostermann, and S. F. Yelin, Dynamic population of multiexcitation subradiant states in incoherently excited atomic arrays, Phys. Rev. A {\bf 107}, L051701 (2023).


\bibitem{DeVoe1996}
R. G. DeVoe and R. G. Brewer, Observation of Superradiant and Subradiant Spontaneous Emission of Two Trapped Ions, Phys. Rev. Lett. {\bf 76}, 2049 (1996).

\bibitem{Bloch2008}
I. Bloch, J. Dalibard, and W. Zwerger, Many-body physics with ultracold gases, Rev. Mod. Phys. {\bf 80}, 885-964 (2008).

\bibitem{Mlynek2014}
J. A. Mlynek et al., Observation of Dicke superradiance for two artificial atoms in a cavity with high decay rate,
Nat. Commun. {\bf 5}, 5186 (2014). 

\bibitem{Guerin2016}
W. Guerin, M. O. Ara\'{u}jo, and R. Kaiser, Subradiance in a Large Cloud of Cold Atoms, Phys. Rev. Lett. {\bf 116}, 083601 (2016).

\bibitem{Jenkins2017}
S. D. Jenkins, J. Ruostekoski, N. Papasimakis, S. Savo, and N. I. Zheludev, Many-Body Subradiant Excitations in Metamaterial Arrays: Experiment and Theory, Phys. Rev. Lett. {\bf 119}, 053901 (2017).

\bibitem{Solano2017}
P. Solano, P. Barberis-Blostein, F. K. Fatemi, L. A. Orozco, and S. L. Rolston, Super- and Sub-Radiance Reveal Infinite-Range Interactions through a Nanofiber, Nat. Commun. {\bf 8}, 1857 (2017).

\bibitem{Rui2020}
J. Rui, D. Wei, A. Rubio-Abadal et al., A subradiant optical mirror formed by a single structured atomic layer, Nature {\bf 583}, 369-374 (2020).

\bibitem{Zhou2024}
X. Zhou, H. Tamura, T. H. Chang, and C. L. Hung, Trapped Atoms and Superradiance on an Integrated Nanophotonic Microring Circuit, Phys. Rev. X {\bf 14}, 031004 (2024).


\bibitem{Rastogi2022}
A. Rastogi et al., Superradiance-Mediated Photon Storage for Broadband Quantum Memory, Phys. Rev. Lett. {\bf 129}, 120502 (2022).

\bibitem{Bonanno2022}
S. Buckley-Bonanno et al., Optimized geometries for cooperative photon storage in an impurity coupled to a two-dimensional atomic array,
Phys. Rev. A {\bf 106}, 053706 (2022).


\bibitem{Facchinetti2016}
G. Facchinetti, S. D. Jenkins, and J. Ruostekoski, Storing light with subradiant correlations in arrays of atoms, Phys. Rev. Lett. {\bf 117}, 243601 (2016).

\bibitem{Ballantine2021}
K. E. Ballantine and J. Ruostekoski, Quantum single-photon control, storage, and entanglement generation with planar atomic arrays, PRX Quantum {\bf 2}, 040362 (2021).

\bibitem{Rubies-Bigorda2022}
O. Rubies-Bigorda, V. Walther, T. L. Patti, and S. F. Yelin, Photon control and coherent interactions via lattice dark states in atomic arrays, Phys. Rev. Res. {\bf 4}, 013110 (2022).


\bibitem{Brody2014}
D. C. Brody, Biorthogonal quantum mechanics, J. Phys. A: Math. Theor. {\bf 47}, 035305 (2014).





\bibitem{Zhang2021}
S. Zhang, P. Ramchurn, K. Bongs, and Y. Singh, Simulation of dipole-dipole interactions with ultracold Sr in an optical lattice, in 2021 Joint Conference of the European Frequency and Time Forum and IEEE International Frequency Control Symposium (EFTF/IFCS) (IEEE, Piscataway, NJ, 2021), pp. 1-3.

\bibitem{Ban2005}
H. Y. Ban, M. Jacka, J. L. Hanssen, J. Reader, and J. J. McClelland, Laser cooling transitions in atomic erbium, Opt. Express {\bf 13}, 3185 (2005).

\bibitem{Kraft1988}
D. Kraft, A software package for sequential quadratic programming, Tech. Rep. DFVLR-FB 88-28, DLR German Aerospace Center, Germany (1988).
















\end{thebibliography}

\end{document}